\documentclass[%
 reprint,
 amsmath,amssymb,
 aps,
]{revtex4-1}
\usepackage{xcolor}
\usepackage{url}
\usepackage{graphicx}
\usepackage{dcolumn}
\usepackage{bm}
\usepackage{amsmath}
\usepackage{epsfig}

\begin{document}

\preprint{APS/123-QED}

\title{Memory-free dynamics for the TAP equations of Ising models \\with arbitrary rotation invariant 
ensembles of random coupling matrices} 

\author{Burak \c{C}akmak and Manfred Opper} \thanks{Both authors are co-first authors.}
\affiliation{Artificial Intelligence Group, Technische Universit\"at Berlin, Germany}%

\date{\today}
\def\mathlette#1#2{{\mathchoice{\mbox{#1$\displaystyle #2$}}%
		{\mbox{#1$\textstyle #2$}}%
		{\mbox{#1$\scriptstyle #2$}}%
		{\mbox{#1$\scriptscriptstyle #2$}}}}
\newcommand{\matr}[1]{\mathlette{\boldmath}{#1}}
\newcommand{\RR}{\mathbb{R}}
\newcommand{\CC}{\mathbb{C}}
\newcommand{\NN}{\mathbb{N}}
\newcommand{\ZZ}{\mathbb{Z}}
\newcommand{\bfl}[1]{{\color{blue}#1}}

\newtheorem{assumption}{Assumption}
\newtheorem{theorem}{Theorem}
\newtheorem{remark}{Remark}
\newtheorem{derivation}{Derivation}
\newtheorem{lemma}{Lemma}
\newtheorem{definition}{Definition}
\newcommand{\new}{\color{red}}
\begin{abstract}
We propose an iterative algorithm for solving the Thouless-Anderson-Palmer (TAP) equations of Ising models with arbitrary rotation invariant (random) coupling matrices. In the thermodynamic limit, we prove by means of the dynamical functional method that the proposed algorithm converges when the so-called de Almeida Thouless (AT) criterion is fulfilled. Moreover, we give exact 
analytical expressions for the rate of the convergence.
\end{abstract}

\maketitle

\section{Introduction}
TAP equations are sets of generalized mean field equations for computing approximate marginal moments in large probabilistic models with a dense connectivity of couplings with random strengths. Originally developed to compute the magnetisations of the Sherrington Kirkpatrick (SK) model of a spin-glass \cite{TAP}, there are now various generalizations especially to models in  the area of statistical data science~\cite{OW0,Kabashima,Donoha,Tramel18}. 

Hence, it is an important to design and study the performance of iterative algorithms which can generate solutions to TAP equations. Several results for the dynamical properties of such algorithms have been derived for models with specific probability distributions of the couplings \cite{Bolthausen,Bayati,bayati2015universality}. More recent research has attempted to extend such results to more general (and possibly more realistic) classes of probability distributions \cite{Opper16,VAMP,takeuchi2017rigorous,Burak17,fletcher2018inference}.

In this paper we will address the dynamics of algorithms for the original case of Ising spin models. The model in its simplest formulation (with constant external fields) discussed in this paper
is less interesting for data science.  But its generalizations have been applied e.g. to modeling the dependencies of spikes recorded from ensembles of neurons \cite{schneidman2006weak,roudi2009ising}, to protein structure determination \cite{weigt2009identification} and gene expression analysis \cite{lezon2006using}. For a review, see \cite{nguyen2017inverse}. In any case, we expect that a proper understanding of algorithms for the simplest setting will also be highly useful for treating more complex models. We will consider TAP equations for the large class of rotation invariant random matrices \cite{Parisi}, which allow for nontrivial dependencies between matrix elements. We will show in Section~\ref{modelpr} that simpler cases such as the independent Gaussian couplings of the SK model or those of the {\em Hopfield model}  and its generalizations \cite{mezard2017mean} are special cases of the rotation invariant ensembles studied in this work.

The analysis of algorithms for TAP equations can be described in terms of the nonlinear dynamics of a system of nodes coupled by a dense random matrix.
In the thermodynamic limit of large systems, it is possible to derive an exact decoupling of the degrees of freedom for the statistical
properties of the dynamics using the so-called {\em dynamical functional} (DF) method \cite{Martin}. Unfortunately, the effective stochastic process of single variables contains couplings between the same variables at different times. This usually precludes closed form analytical results for the dynamical properties. In a previous paper,
motivated by existing algorithms for the case of specific random matrices, we have shown that such temporal self couplings can be canceled for the entire class of rotation invariant random matrices, if one includes specific memory terms in the algorithm. In that case,
the stochastic fields which drive the effective dynamics reduce to a Gaussian process.  

Our previous construction of the so-called {\em single-step memory} (SSM) algorithm for solving TAP equations presented in \cite{Opper16} comes with two major drawbacks. First, the memory terms in the algorithm are based on the coefficients of 
the power series expansion of a specific function derived
from the random matrix ensemble. Of course, these can be obtained easily, when an analytical 
expression of the function is known. But for any given large matrix which is 
assumed to be drawn at random from a rotation invariant (but otherwise 
unknown) ensemble it is not clear 
how to obtain such coefficients numerically in a simple way. This problem will make the algorithm not a good candidate for practical applications.
Second, from a theoretical perspective our analytical analysis is still incomplete. We were able to show by a linear stability analysis that the well-known {\em de Almeida Thouless} (AT) \cite{AT}
criterion which separates an ergodic region of the spin-model from 
a more complex phase, is necessary for convergence of the algorithm. Unfortunately, we were
not able to get general analytical results for the asymptotic convergence rates. 

In this paper we present a new construction for an algorithm for the Ising system which was motivated by similar approaches in the field of compressed sensing \cite{VAMP,takeuchi2017rigorous}. Remarkably, the updates of the algorithm
avoid the use of explicit memory terms and 
can be easily applied to an Ising model with a given matrix of couplings.
The analysis by the DF method again shows a vanishing of the self-couplings.
This leaves us with an effective dynamics that is driven by a Gaussian process
but with a much simpler covariance structure compared to our previous method. 
We are able to prove convergence  from a random initialization 
of the algorithm in the thermodynamic limit provided that the AT criterion is fulfilled. 
Moreover, we also obtain explicit expressions for the rate of convergence.
To our knowledge this is the first time that analytical results for the convergence of
such algorithms have been obtained.

The paper is organized as follows: In Section~2 we introduce the model and present motivating 
examples for studying rotation invariant ensembles of coupling matrices. Section~3 provides a brief presentation on the TAP equations. In Section~4 we present our new algorithm for solving the TAP equations and in Section~5 we study its properties in the thermodynamic limit. Section~6 provides convergence properties of the algorithm. In section~7 we show how to compute 
parameters which are needed by the algorithm before the iteration starts. For comparison, 
in Section~8 we
give analytical expressions for the convergence rates of 
the SSM algorithm of \cite{Opper16} for the special cases of the SK and Hopfield models. Comparisons of the theory with simulations are given in Section~9. Section~10 presents a summary and outlook. Major derivations of our results are located at the Appendix.

\section{Ising models with random couplings}
We consider Ising models with pairwise interactions of the spins $\matr s=(s_1,\ldots,s_N)^\top\in\{-1,1\}^{N}$ described by the (conditional) Gibbs distribution 
\begin{equation}
p(\matr s\vert \matr J,\matr h)\doteq \frac{1}{Z}\exp\left(\frac{1}{2}\matr s^\top\matr J\matr s+\matr s^\top\matr h\right)\label{Gibbs}
\end{equation}
where $Z$ stands for the normalization constant. Our concern is to compute the vector of magnetizations 
\begin{equation}
\matr {m}=\mathbb E[\matr s]
\end{equation}
where the expectation is taken over the Gibbs distribution. For the sake of simplicity of the analysis we limit our attention to the case where all external fields are equal
\begin{equation}
h_{i}=h\neq 0,~\forall i.
\end{equation}
 
We assume that the coupling matrix $\matr J=\matr J^{\top}$ is drawn from a rotation invariant matrix ensemble, i.e. $\matr J$ and $\matr V\matr J\matr V^\top$ have the same probability distributions for any orthogonal matrix $\matr V$ independent of $\matr J$. Equivalently, $\matr J$ has the spectral decomposition \cite{Collins14}
\begin{equation}
\matr J=\matr O^
\top\matr D\matr O \label{decom}
\end{equation}
where $\matr O$ is a Haar (orthogonal) matrix that is independent of a real-diagonal matrix $\matr D$. Though our analysis does not make reference to any specific coupling matrix model, 
it is interesting to note, that certain models derived from products of simpler matrices
which were originally introduced in the context of communication theory \cite{ralfa} and recently reappeared in the statistical physics context \cite{mezard2017mean}, belong to the rotation invariant family.
\subsection{Motivation: Models with a product of random matrices}\label{modelpr}
Consider the product of $M$ matrices  
\begin{equation}
\matr X\doteq\matr X_M\matr X_{M-1}\cdots \matr X_1 \label{prod}
\end{equation}
where $\matr X_m\in\RR^{N_{m}\times N_{m-1}}$ and $N_0\doteq N$. We will consider
coupling matrices of the form  
\begin{equation}
\matr {J}=\beta\matr X^\top\matr X-\frac{\beta}{N}{\rm tr}(\matr X^\top\matr X){\bf I}\label{Mlayer}
\end{equation}
where $\beta$ stands for the inverse temperature parameter. With this modeling of the coupling matrix the Ising model becomes equivalent to an M-layer probabilistic feed-forward neural network
with linear transfer functions. Specifically, for a set of hidden node vectors  $\mathcal U\doteq\{\matr u_0,\matr u_1,\cdots,\matr u_M\}$ with $\matr u_m\in \RR ^{N_m\times 1}$ we define an $M$-layer linear neural network by the distribution 
\begin{align}
	&p(\matr s,\mathcal U\vert \mathcal X,\matr h)\doteq\frac{1}{Z}e^{-\frac{\tilde\beta}{2}+\matr s^\top(\matr u_0+\matr h)}\times \nonumber \\ &\quad \times \left(\prod_{m=1}^{M}\delta(\matr u_{m-1}-\matr X_m^\top\matr u_m)\right){N}\left(\matr u_M;\matr 0,\frac 1 \beta {\bf I}\right)
\end{align}
where $\mathcal X\doteq \{\matr X_1,\ldots,\matr X_M\}$, $\tilde\beta\doteq\beta{\rm tr}(\matr X^\top\matr X)$ and ${N} (\cdot; \matr \mu,\matr \Sigma)$ denotes the Gaussian density function with mean $\matr \mu$ and covariance $\matr \Sigma$. It is easy to see that by integrating out the Gaussian variables $\mathcal U$, we arrive at the Gibbs distribution (\ref{Gibbs})
\begin{equation}
p(\matr s\vert \matr J,\matr h)=\int {\rm d}\mathcal U\; p(\matr s,\mathcal U\vert \mathcal X, \matr h).
\end{equation}
  
We will next discuss some interesting cases for which this model leads to rotation invariant ensembles.
	\subsubsection{Multi-layer Hopfield model}	 
Let $\matr J$ be as in \eqref{Mlayer} and let all $N_m\times N_{m-1}$ matrices $\matr X_m$ in \eqref{prod} be independent where the entries of $\matr X_m$ are independent Gaussian 
random variables with zero mean and variance $1/N_{m}$ \cite{ralfa}. The case $M=1$ 
is essentially the well-known {\em Hopfield model} \cite{Hopfield}. Here, we ignore the fact that in the present case, the stored ``patterns'' are Gaussian not Ising variables. 
Since we are not interested in memory retrieval properties of the model, the difference is not relevant. 

The case $M=2$ coincides with Mezard's {\em combinatorial disorder Hopfield model} \cite{Mezard}. In general, random matrices $\matr J$ constructed in such a way belong
to rotation invariant ensembles and we call the model {\em multi-layer Hopfield model}. 
\subsubsection{Multi-layer random orthogonal model}\label{mrom}
Let $\matr J$ be as in \eqref{Mlayer} and the $N_m\times N_{m-1}$ matrices $\matr X_m$ in \eqref{prod} are defined as
\begin{equation}
\matr X_m=\frac{1}{\sqrt{\rho_{m}}}\matr P^L_{\rho_{m}}\matr O_m(\matr P^L_{\rho_{m-1}})^\top\label{rom}
\end{equation}
where all $\matr {O}_m$ are independent $L\times L$ Haar matrices, and $\matr P^{L}_{\rho}$ is a $\rho L\times L$ matrix with $\rho\leq 1$ and 
\begin{equation}
(\matr P^{L}_{\rho})_{ij}=\delta_{ij},\quad \forall i,j.
\end{equation}
For $M=1$ and $\rho_0=1$ we arrive at the {\em random orthogonal model} of Parisi and Potters \cite{Parisi}. In general, matrices $\matr {J}$ are also rotation invariant and we call the model \emph{multi-layer random orthogonal model}.

\subsubsection{SK model as the universal small $\alpha$-limit}
We next consider the scaling properties of the coupling matrix model \eqref{Mlayer} when the aspect ratio $\alpha\doteq \frac{N_0}{N_M}$  of the $N_M\times N_0$ matrix $\matr X$ in \eqref{prod} becomes small. For the 1-layer Hopfield model it is well-known that by rescaling
$\matr J\to \frac{1}{\sqrt{\alpha}}\matr J$
of the coupling matrix the model becomes equivalent to the SK model \cite{mezard2017mean} in the limit $\alpha\to 0$. To be precise, the limiting spectral distribution of $\frac{1}{\sqrt{\alpha}}\matr J$ becomes a Wigner semicircle law in the limit $\alpha\to 0$. We next show that this small-$\alpha$ limit holds in a universal sense for both the multi-layer Hopfield and multi-layer random orthogonal models. Specifically, the coupling matrices for $\alpha\leq 1$ are in general of the form 
\begin{equation}
\matr J=\matr P^{N_M}_\alpha\matr Y(\matr P^{N_M}_\alpha)^\top.\label{skform}
\end{equation}
Let $f_{\alpha}$ denote the density of the limiting spectral distribution of $\frac{1}{\sqrt{\alpha}}\matr J$ and let $\sigma_{\matr Y}^2\doteq\lim_{N\to \infty}\frac{1}{N}{\rm tr}(\matr Y^2)$. Then,
we show in Appendix~\ref{dlimitsk} that
\begin{equation}
\lim_{\alpha\to 0}{f}_\alpha(x)=\frac {{\sqrt{4\sigma_{\matr Y}^2-x^2}}}{2\pi \sigma_{\matr Y}^2} \;, \quad \vert x \vert \leq 2\sigma_{\matr Y}.\label{limitsk}
\end{equation}
Actually, this result holds for any coupling matrix of the form \eqref{skform} with rotation invariant $\matr Y$ which has a compactly supported limiting spectral distribution with $\lim_{N\to \infty}\frac{1}{N}{\rm tr}(\matr Y)=0$. We note that $\sigma_{\matr Y}^2$ is a model-dependent parameter, e.g. for the multi-layer Hopfield model we have 
\begin{equation}
\sigma_{\matr Y}^2=\beta^2(1+\sum_{m<M}\alpha_m)
\end{equation}
with $\alpha_m\doteq N_0/N_m$.  For a convenient computation of the model parameter $\sigma_{\matr Y}^2$ in general we refer the interested reader to Appendix~\ref{addsigma}. 
\section{TAP equations}
The TAP equations are a set of nonlinear equations for approximating the magnetizations $\matr m$ of the Ising model with random couplings.
They are assumed (under certain conditions) to give the exact results in the thermodynamic limit \cite{Mezard} and were first proposed for the SK model in \cite{SK}. A generalization to rotation invariant matrices were originally derived in \cite{Parisi} using 
a free energy approach. An alternative derivation was given in \cite{Adatap} using the cavity method. We also refer  to \cite{CakmakOpper18} for a rigorous approach on the self averaging assumptions made for cavity field variances in this approach. The TAP equations are given by
\begin{subequations}
\label{tap}
\begin{align}
\matr m&={\rm Th}(\matr \gamma)\\
\matr \gamma&=\matr J\matr m-{\rm R}(\chi)\matr m\\
\chi &= \mathbb E_u[{\rm Th}'(\sqrt{(1-\chi){\rm R}'(\chi)} u)] \label{chi}.
\end{align} 
\end{subequations}
Here, for convenience we define the function 
$${\rm Th}(x)\doteq\tanh(h+x) .$$ The random variable $u$ is a standard (zero mean, unit variance) normal Gaussian. 

The only dependency on the random matrix ensemble of $\matr J$ is through the
R-transform which is defined as \cite{mingo2017free}
\begin{equation}
{\rm R}(\omega)={\rm G}^{-1}(\omega)-\frac{1}{\omega},  \label{Rtrans}
\end{equation}
where ${\rm G}^{-1}$ is the functional inverse of the Green-function 
\begin{equation}
{\rm G}(z)\doteq \lim_{N\to\infty}\frac{1}{N}{\rm tr}((z{\bf I}-\matr J)^{-1}) . \label{Greens}
\end{equation}
${\rm G}(z)$ is bijective for real $z$ which are not in the support of the limiting spectral distribution of~$\matr J$. Hence, ${\rm R}(\omega)$ is well-defined for ${\rm G}_{\min}<\omega< {\rm G}_{\max}$ where we have defined ${\rm G}_{\max}\doteq\lim_{z\to \lambda_{\max}}{\rm G}(z)$
(and similarly for ${\rm G}_{\min}$) with $\lambda_{\max,\min}$ denoting the supremum and infimum of the support, respectively. One can show that
${\rm R}(\omega)$ is strictly increasing \cite{Guionnet} (unless $\matr J=c{\bf I}$ for some $c\in \RR$), i.e. ${\rm R}'(\omega)>0$  where ${\rm R}'$ stands for the derivative of ${\rm R}$.

As an example, for the (1-layer) Hopfield model with  $\alpha_1\doteq N_0/N_1$
we have \cite{Opper16}
\begin{equation}
{{\rm R}({\omega})}=\frac{\alpha_1\beta^2\omega}{1-\alpha_1\beta\omega}.\label{R1}
\end{equation}
The corresponding equations \eqref{tap} agree with the well-known TAP equations of the Hopfield model previously obtained e.g. 
in \cite{Mezard}. {Moreover,} for the 2-layer Hopfield model with $\alpha_2\doteq N_0/N_2$ we have (see Appendix~\ref{drmodel2})
\begin{equation}
{\rm R}({\omega})=\frac{\frac{1}{\beta \omega}-\alpha_1-\alpha_2-\sqrt{(\frac{1}{\beta \omega}-\alpha_1-\alpha_2)^2-4\alpha_1\alpha_2}}{2\alpha_1\alpha_2\omega}-\beta\label{rmodel2}
\end{equation}
and the resulting TAP equations \eqref{tap} coincide with those derived in \cite{mezard2017mean} by means of a mean field message passing method. 

The TAP equations \eqref{tap} are well-defined when
\begin{align}
\eta{\rm R}'(\chi)&< 1 \label{cond2}~~ ({\rm AT})\\
\chi&<{\rm G}_{\max} \label{cond1}	
\end{align}
where we have set
\begin{equation}
\eta\doteq\mathbb E_u[({\rm Th}'(\sqrt{(1-\chi){\rm R}'(\chi)} u))^2].\label{eta}
\end{equation}
The condition (AT) is sufficient for $\chi$ to have a unique solution in \eqref{chi}. This result can be easily derived by following the arguments of the derivation of \cite[Lemma 2.2]{Bolthausen}. There, the result refers to the SK model, i.e. ${\rm R}(\omega)=\beta^2\omega$.
Equality in \eqref{cond2}, i.e. $\eta{\rm R}'(\chi)=1$, agrees with the well-known AT 
line of stability \cite{AT} for the Replica-symmetric ansatz for spin models with rotation invariant coupling matrices, see \cite[Eq. (46)]{Enzooo}. In the region of stability, we can also expect that the solution of the TAP equations will give us the asymptotically correct magnetisations in the thermodynamic limit. 
The second condition \eqref{cond1} is needed to ensure that the Green-function has a unique
inverse, such that the R-transform can be defined properly. For the SK-model, e.g.
it restricts the valid inverse temperature regions to $\beta<\frac{1}{\chi}$. Similarly, for the 
(1-layer) Hopfield model, one needs  $\beta<\frac 1 \chi\frac{\alpha_1}{1+\sqrt{\alpha_1}}$. 

\section{Iterative solution of the TAP equations}
In this section we define a new iterative algorithm for solving the TAP equations \eqref{tap}. First, we recall that the only dependency on the random matrix ensemble in the TAP equations \eqref{tap} is via the R-transform ${\rm R}\doteq{\rm R}(\chi)$ (for short) and its derivative ${\rm R}'\doteq {\rm R}'(\chi)$.
Hence, if the limiting spectral distribution of the random matrix ensemble is known, these quantities can be computed exactly from the
Green-function (\ref{Greens}) via (\ref{Rtrans}). We will show in {Section~\ref{algorc}} a practical computation
of these quantities for concrete realizations of matrices $\matr J$.

We are looking for a solution to the TAP equations in terms of an iteration of a vector of auxiliary variables
$\matr \gamma(t)$, where $t$ denotes the discrete time index of the iteration.
The initialization of this vector is given by $\matr \gamma(0)=\sqrt{(1-\chi){\rm R}'} \matr u$ where $\matr u$ is a vector of independent normal Gaussian random variables. 
We then proceed by iterating 
\begin{subequations}
\label{dynamics}	
	\begin{align}
		\tilde{\matr \gamma}(t)&=\frac{1}{\chi}{\rm Th}(\matr \gamma(t-1))-\matr \gamma(t-1)\\
		\matr \gamma(t)&=\matr A\tilde{\matr \gamma}(t) \label{TAP_dyn_two}
	\end{align} 
\end{subequations}
for $t=1,2, 3, \ldots,T$ with the \emph{time-independent} random matrix
\begin{equation}
\matr A\doteq\frac{1}{\chi}(\lambda{\bf I}-\matr J)^{-1} -\bf I.
\label{def_matrixA}
\end{equation}  
The variable $\lambda$ is defined as the (unique) solution of the scalar equation
\begin{equation}
{\rm G}(\lambda) = \chi
\end{equation}  
or equivalent (see (\ref{Rtrans}))
\begin{equation}
{\rm R}(\chi) = \lambda - \frac{1}{\chi} \label{lamdef} .
\end{equation}  
It is easy to see that the fixed points of $\matr \gamma(t)$ coincide with 
the solution of the TAP equations for $\matr\gamma$, \eqref{tap}, 
if we identify the corresponding magnetisations by
\begin{equation}
\matr m = \chi(\tilde{\matr \gamma} +\matr \gamma).\label{magnez}
\end{equation}  
\section{Dynamics in the thermodynamic limit}
Our goal is to analyze the convergence properties of the sequence $\matr \gamma(t)$  
in the thermodynamic limit by studying 
the deviation between the dynamical variables at different times 
\begin{align}
	\Delta(t,s)\doteq \lim_{N\to\infty}\frac{1}{N} \mathbb E[\Vert \matr \gamma(t)-\matr \gamma(s) \Vert^2]
	\label{measure}
\end{align}
where the expectation is taken over the random matrix ensemble of $\matr A$ and the random initialization $\matr \gamma(0)$.
In the thermodynamic limit, the normalized squared distance will become self averaging and represents the typical squared distance corresponding to a single realization of the dynamics \eqref{dynamics}. We will show
later, that as $t,s\to\infty$, the deviation $\Delta(t,s)$ will converge to $0$
showing that the sequence of iterates $\matr \gamma(t)$ will have a limit which 
thus solves the TAP equations.

Dynamical properties of fully connected disordered systems can be computed by a discrete time version of the method of DF developed by Martin, Siggia and Rose \cite{Martin}
and also used extensively for studying spin-glass dynamics, see e.g. \cite{sompo82}, \cite{Eisfeller}. This method provides us with exact results for the thermodynamic properties of the marginal distribution of a trajectory of arbitrary length $T$, for {\em single variables} (in our case $\gamma_i(t)$ and $\tilde\gamma_i(t)$) as long as the large-system limit $N\to\infty$ is taken before the long-time limit $T\to\infty$. The method is based on averaging the moment \emph{generating functional} for the trajectory of $\gamma_i(t)$ and $\tilde\gamma_i(t)$
\begin{align}
Z\{l_i(t)\}\doteq\int \prod_{t=1}^T &{\rm d}\tilde{\matr\gamma}(t) {\rm d}\matr \gamma(t)\; \delta\left(\matr{\tilde\gamma}(t)-{f}(\matr \gamma(t-1))\right)\times\nonumber \\&\times\delta\left(\matr \gamma(t)-\matr A\matr{\tilde\gamma}(t)\right)e^{{\rm i}\gamma_i(t) l_i(t)}
\end{align}
over the random matrix $\matr A$ and the random initialization $\matr \gamma(0)$. Here, for short we introduce the function 
\begin{equation}
f(x)\doteq \frac{1}{\chi}{\rm Th}(x)-x.
\end{equation}
The generating functional obviously contains the correct dynamics \eqref{dynamics} within the Dirac delta $\delta$ functions. Moreover, since the partition function is normalized, i.e. $Z\{l_i(t)=0\}=1$, the replica-method is \emph{not needed} to perform the expectations.

By construction, the distribution of $\matr A$ induced by the matrix $\matr J$ is also rotation invariant. Hence we can follow the steps of a previous paper \cite[Appendix C]{Opper16}, by
essentially replacing all averages over the matrix $\matr J$ in that paper by averages over $\matr A$
and read off the result.
In our case, $\tilde{\matr\gamma}(t)$ plays the role of $\matr m(t)$~in~\cite{Opper16}.
We then find that the averaged generating functional 
$\mathbb E [Z(\{l_i(t)\})]$ becomes
\begin{align}
& \int{\rm d}\mathcal N(\{\phi (t)\};\matr 0,\mathcal C_\phi) \
{\rm d}\mathcal N(\gamma_i(0);0, (1-\chi){\rm R}') 
\nonumber \times \\
&\times\;\prod_{t=1}^{T}{\rm d}\tilde\gamma_i(t){\rm d}\gamma_i(t)\;\delta( \tilde\gamma_i(t)- f(\gamma_i(t-1)))\nonumber \\
&\times\delta\left(\gamma_i(t)-\sum_{s< t}\mathcal{\hat G}(t,s) {\tilde \gamma_i}(s)-\phi(t)\right) e^{{\rm i}\gamma_i(t)l_i(t)+\epsilon} \label{dftresult}
\end{align}
where $\epsilon=O(1/N)$ is a constant term that, as indicated, vanishes as $N \to \infty$ and $\mathcal N({\cdot};\matr \mu, \matr \Sigma)$ denotes the Gaussian distribution function with mean $\matr \mu$ and covariance~$\matr \Sigma$. Hence, through averaging, we manage to \emph{decouple}
the component $\gamma_i(t)$  in \eqref{dynamics} from the remaining components
for $N\to\infty$ and we obtain an \emph{effective stochastic process} for the dynamics of single, arbitrary components $\gamma(t)$ and $\tilde\gamma(t)$ (where we omit the index $i$ in the following)
of the vectors $\matr\gamma(t)$ and $\tilde{\matr\gamma}(t)$, respectively. This contains all information about the marginal statistics of these variables 
induced by the randomness of $\matr A$ and $\matr \gamma (0)$. Note that ${\rm d}\mathcal N(\{\phi (t)\};{ \matr 0},\mathcal C_\phi)$
represents an average over a discrete time Gaussian process $\phi(t)$ with zero mean 
and covariance $\mathcal C_\phi$.

The resulting effective stochastic process of single variables is immediately read off
from (\ref{dftresult}) as
\begin{subequations}
	\label{esp}
	\begin{align}
	\tilde\gamma(t)&=f(\gamma(t-1))\label{gam1}\\
	\gamma(t)&=  \sum_{s < t}{\hat {\mathcal G}(t,s)}\tilde\gamma(s) + \phi(t) \label{gam2}
	\end{align}
\end{subequations}
for $t=1,2,3\ldots$ with the initial condition $\gamma(0) = \sqrt{(1-\chi){\rm R}'} u$
where $u$ is standard normal Gaussian. For a trajectory of length $T$, ${\hat {\mathcal G}(t,s)}$ are given by the $(t,s)$ indexed entries of the $T\times T$ matrix ${\hat {\mathcal G}}$
which is defined by the matrix function 
\begin{equation}
\hat{\mathcal G}\doteq{\rm R}_{\matr A}(\mathcal G) \label{Ghat}
\end{equation} 
where ${\rm R}_{\matr A}$ is the R-transform corresponding to the random matrix
${\matr A}$ (see the definitions in eq (\ref{Rtrans})).
$\mathcal G$ is a $T\times T$ order parameter matrix, which must be computed from the
entire ensemble of trajectories of  $\gamma(t)$. Its
$(t,s)$ indexed entries are defined by the response function
\begin{equation}
\mathcal G(t,s)\doteq\mathbb E\left[\frac{\partial\tilde\gamma(t)}{\partial \phi (s)}\right]  .
\label{resp}
\end{equation}
By causality $\mathcal G$ is an upper triangular matrix (i.e. its $(t,s)$ indexed entries are zero for $s \geq t$).
The R-transform of a matrix can be defined from the power series expansion of the R-transform \cite{mingo2017free}
\begin{equation}
{\rm R}_{\matr A}(\omega)\doteq\sum_{n=1}^{\infty}c_{\matr A,n} \omega^{n-1} \label{R_series}
\end{equation}
in terms of the so-called {\emph  free cumulants} $c_{\matr A,n}$
by inserting the powers of the matrix $\mathcal G^{n-1}$ for $\omega^{n-1}$.
It follows that $\hat{\mathcal G}$ is
also upper triangular.  From the definition of the matrix $\matr A$, (\ref{def_matrixA}) and standard properties of R-transform, one can show that 
\begin{align}
c_{\matr A,1} &= \lim_{N\to\infty} \frac{1}{N} {\rm tr} (\matr A) = 0\\
c_{\matr A,2} & = \lim_{N\to\infty} \frac{1}{N} {\rm tr} (\matr A^2) - c^2_{\matr A,1}  \\
& = \lim_{N\to\infty}
\left(\frac{1}{\chi^{2}}\frac{1}{N} {\rm tr}((\lambda {\bf I}-\matr J)^{-2})-1\right) \label{sline}\\
& =
\frac{\chi ^2{\rm R}'}{1-\chi^2{\rm R}'}
\label{c_results}
\end{align}
where in \eqref{sline} we use ${\rm G}'(\lambda) =-\lim_{N\to\infty} \frac{1}{N} {\rm tr}((\lambda {\bf I}-\matr J)^{-2})$ and re-express the result in terms of the R-transform of the limiting spectral distribution of $\matr J$.

Finally, the zero-mean Gaussian process $\{\phi(t)\}$ has a 
covariance matrix given by
\begin{equation}
\mathcal C_\phi = \sum_{n=1}^{\infty}c_{\matr A,n}\sum_{k=0}^{n-2} \mathcal {G}^k\mathcal {\tilde C}(\mathcal {G}^\top)^{n-2-k} \label{cov}
\end{equation}
where 
\begin{equation}
\mathcal {\tilde C}(t,s)\doteq \mathbb E[\tilde\gamma(t)\tilde \gamma(s)] 
\label{tilde_cov}.
\end{equation}
and the $c_{\matr A,n}$ are defined by the R-transform (\ref{R_series}).

It may seem that we have not gained much from this approach, because interactions between different dynamical variables are now replaced by couplings of single variables over time.
In principle, all matrix elements of  $\hat{\mathcal G}$, $\mathcal G$ and $\mathcal C$ could be (approximately) computed sequentially in time, e.g. by lengthy Monte Carlo simulations of the process
$\phi(t)$ \cite{Eisfeller}. On the other hand, the memory terms preclude an analytical treatment in general. Surprisingly, as we shall show in Appendix \ref{vanishmemory} the memory terms vanish, i.e. we get
\begin{equation}
\hat{\mathcal G}=\mathcal{G}= {\matr 0}.\label{memoryless}
\end{equation}
The memory-free property \eqref{memoryless} implies that 
\begin{equation}
\gamma (t)=\phi(t) \label{phi_gamma}
\end{equation}
for $t=1,2,\ldots,T$ are Gaussian random variables with a much simplified covariance structure that is given by
\begin{align}
{\mathcal C}&= c_{\matr A,2}\tilde{\mathcal C}\\
&=\frac{\chi ^2{\rm R}'}{1-\chi^2{\rm R}'}\tilde{\mathcal C}.
\end{align}
Explicit results can be computed by the recursion (see Appendix~\ref{dcovfield})
\begin{subequations}
	\label{covfield}
\begin{align}
 \mathcal C(t,s)&=\frac{g(\mathcal C(t-1,s-1))}{1/{\rm R'}-\chi^2}  \label{covf} \\
 \mathcal C(t,t)&= (1-\chi){\rm R}'
\end{align}
\end{subequations} 
where we have defined
\begin{equation}
g(x)\doteq {\mathbb E[{\rm Th}(\gamma_1){\rm Th}(\gamma_2)]} -\chi^2x\label{kfunc}
\end{equation}
such that the expectation is taken over the random variables $\gamma_1$ and $\gamma_2$ which are jointly Gaussian  with zero mean and equal variance $(1-\chi){\rm R}'$ and covariance 
\begin{equation}
x\doteq \mathbb E[\gamma_1\gamma_2].
\end{equation}
To initialize the recursion (\ref{covf}), we note that by (\ref{phi_gamma}) we have 
$\mathcal C(t,s) = \mathbb E[\gamma(t) \gamma(s)]$ and we can include the initialization $\gamma(0)$ in the
definition. It is easy to see that the random variable $\gamma(0)$ is independent 
from $\gamma(t)$ for $t > 0$. Hence, we get
\begin{equation}
\mathcal C(t,0) = 0,\quad t>0\label{s1}.
\end{equation}

\section{Convergence of the single-variable dynamics} 
The convergence properties of the dynamics can now be obtained from the covariance 
using the equality
\begin{align}
\Delta(t,s) & = \mathcal C(t,t)+\mathcal C(s,s)-2\mathcal C(t,s)\label{deltaC} \\
& = 2(1-\chi){\rm R}'-2\mathcal C(t,s) \label{npt}.
\end{align}
From these results, we are able to show that
\begin{align}
	&\Delta(t+1,s+1)<\Delta(t,s),\quad \forall t,s.\label{res1}
	\end{align}
This means that with increasing time, the distance between pairs of iterates
separated by a fixed amount of time strictly decreases. We thus get
\begin{equation}
\lim_{t,s\to \infty} \Delta(t,s)=0\label{conver}
\end{equation}
and the dynamics is convergent.  Let us denote the convergence rate of $\Delta(t,s)$ by
\begin{align}
\mu_{mf}\doteq\lim_{t,s\to \infty}\frac{\Delta(t+1,s+1)}{\Delta(t,s)}
\end{align}
where for convenience we introduce the abbreviation ``\emph{mf}'' in the notation to emphasize the thermodynamic memory-free property \eqref{memoryless}. We obtain
\begin{equation}
\mu_{mf}=1-\frac{1-\eta{\rm R}'}{1-\chi^2{\rm R}'}.\label{res2}
\end{equation}
Moreover, it turns out that $\ln\mu_{mf}$ gives the exponential decay rate:
\begin{equation}
\lim_{t\to \infty}\frac{1}{t}\ln \Delta(t,\infty)=\ln \mu_{mf}\label{res3}
\end{equation}
where $\Delta(t,\infty)\doteq \lim_{s\to\infty}\Delta(t,s)$. Notice that when the model parameters approach the AT-line, i.e. $\eta{\rm R'}=1$, the convergence becomes arbitrarily slow. The derivations of \eqref{res1}, \eqref{res2} and \eqref{res3} are located at Appendix~\ref{dres13}.

\section{Algorithmic considerations}\label{algorc}
So far we have assumed that the limiting spectral distribution of $\matr J$ is analytically known beforehand for computing the quantities ${\rm R}$ and ${\rm R}'$. In the sequel, we present a practical approach that bypasses this need. 

Specifically, we first compute the spectral decomposition 
\begin{equation}
\matr J=\matr O^\top\matr D\matr O.\label{sdcm}
\end{equation}
The Green function and its derivative are then replaced by their finite-size approximations as 
\begin{align}
{\rm G}(z)&= \frac{1}{N} {\rm tr}((z{\bf I}-\matr D)^{-1}) \\
{\rm G}'(z)&= - \frac{1}{N}{\rm tr}((z{\bf I}-\matr D)^{-2}).
\end{align}
The necessary R-transform and its derivative
are then obtained by solving the  simple fixed-point equations
\begin{subequations}
	\begin{align}
		\lambda &={\rm R}+\frac{1}{\chi}\\
		{\rm R}&=\lambda-\frac{1}{{\rm G}(\lambda)}\\
		{\rm R}'&=\frac{1}{{\rm G}(\lambda)^2}+\frac{1}{{\rm G}'(\lambda)}\\
		\chi &= \mathbb E_u[{\rm Th}'(\sqrt{(1-{\rm G}(\lambda)){\rm R}'} u)].		
	\end{align} 
\end{subequations}
The matrix $\matr A$ is then computed as
\begin{equation}
\matr A=\frac{1}{\chi}\matr O^\top(\lambda{\bf I}-\matr D)^{-1}\matr O-\bf I.\label{resolvent}
\end{equation}  
These steps have to be performed before the main iterations and have $O(N^3)$, i.e.
\emph{cubic} computational complexities. On the other hand, the iterations are just based on matrix vector multiplications and evaluations 
of scalar nonlinear functions. These have only \emph{quadratic} computational complexity per iteration. 
\section{Comparison with a previous algorithm in the cases of SK and Hopfield Models}
In a  previous paper \cite{Opper16}, we have introduced the SSM iterative algorithm for solving TAP equations for models with general rotation invariant coupling matrices. This algorithm and its corresponding DF analysis is based on the
coefficients of power series expansions related to R-transforms of the random matrix ensemble. This practically limits the applicability of the method to cases where such coefficients 
are known analytically. Nevertheless, both the algorithm and its analysis become simple for the SK and Hopfield models. Hence, it is interesting to derive the corresponding convergence rates of SSM algorithms for these models which is what we show in the sequel. 

As for the SK model we first note that the necessary R-transform quantities are given by
\begin{equation}
{\rm R}=\beta^2\chi ~~\text{and}~~ {{\rm R}'}=\beta^2
\end{equation}
where $\chi$ is the solution of \eqref{chi}. The SSM dynamics for the SK model is nothing but 
Bolthausen's celebrated convergent dynamics \cite{Bolthausen}:
\begin{subequations}
	\label{amp_sk}
	\begin{align}
		\matr m(t)&={\rm Th}(\matr \rho(t-1))\\
		\matr \rho(t)&=\matr J\matr m(t)-{\rm R}\matr m(t-1)\label{psi_sk}.
	\end{align}
\end{subequations}
Here, $\matr m(0)\doteq \matr 0$ and $\matr \rho(0)\doteq\sqrt{(1-\chi){\rm R}'}\matr u$ where $\matr u$ is a vector of independent normal Gaussian random variables.

Second, for the Hopfield model, the R-transform and its derivative are (see (\ref{R1}))
\begin{equation}
{\rm R}=\frac{\beta^2\alpha_1\chi}{1-\beta\alpha_1\chi} ~~\text{and}~~ {{\rm R}'}=\frac{\beta^2\alpha_1}{(1-\beta\alpha_1\chi)^2}\label{hopfieldR}
\end{equation}
where $\chi$ is the solution of \eqref{chi}. In the spirit of Bolthausen's dynamics \eqref{amp_sk}, we introduce a SSM construction for the Hopfield model as
\begin{subequations}
	\label{amp}
	\begin{align}
	\matr m(t)&={\rm Th}(\matr \rho(t-1))\\
	\matr \phi(t)&=\matr J\matr m(t)-{\rm R}\matr m(t-1)\\
	\matr\rho(t)&=(1+\frac{\rm R}{\beta})\matr \phi(t)-\frac{\rm R}{\beta}\matr\rho(t-1) \label{psi_hop}
	\end{align}	
\end{subequations}
where $\matr m(0)\doteq \matr 0$ and $\matr \rho(0)\doteq\sqrt{(1-\chi){\rm R}'}\matr u$.

Note that the variable $\matr \gamma(t)$ in the dynamics \eqref{dynamics}
has the same fixed points as the variables $\matr \rho(t)$ in the dynamics \eqref{amp_sk} and \eqref{amp} for the SK and Hopfield models, respectively. In order to analyze the convergence properties of the SSM algorithms \eqref{amp_sk} and \eqref{amp} in the thermodynamic limit we define
\begin{equation}
\Delta_{ssm}(t,s)\doteq \lim_{N\to\infty}\frac{1}{N} \mathbb E[\Vert \matr \rho(t)-\matr \rho(s) \Vert^2]. \label{deltaphi}
\end{equation}
For both the SK and Hopfield models, we have
\begin{equation}
	\Delta_{ssm}(t+1,s+1)<\Delta_{ssm}(t,s),\quad \forall t,s\label{delta_ssm}.
\end{equation} 
Thus, $\Delta_{ssm}(t,s)$ converges to $0$ as $t,s\to\infty$. In particular, the rate of convergence is obtained as
\begin{align}
	\mu_{ssm} & \doteq \lim_{t,s\to \infty}\frac{\Delta_{ssm}(t+1,s+1)}{\Delta_{ssm}(t,s)}=\eta{\rm R}'\label{rho_con}.
\end{align}
The derivations of \eqref{delta_ssm} and \eqref{rho_con} are given in Appendix~\ref{amp_con}.

From a computational point of view, for the cases of the SK model and the Hopfield model, the SSM algorithms  \eqref{amp_sk}  and \eqref{amp} are more convenient than our new algorithm \eqref{dynamics}, because
 the latter requires the computation of a matrix inverse \eqref{def_matrixA} (before the iteration starts). Nevertheless, it is interesting to note from \eqref{res2} that
\begin{equation}
\mu_{mf}-\mu_{ssm}=\frac{{\chi^2{\rm R}'}(\eta{\rm R}'-1)}{1-\chi^2{\rm R}'}<0\label{iden}.
\end{equation}
Thus, in the thermodynamic limit the SSM algorithms for SK and Hopfield models have both a slower convergence compare to the algorithm \eqref{dynamics}. 

\section{Simulation Results}
\begin{figure}
	\epsfig{file=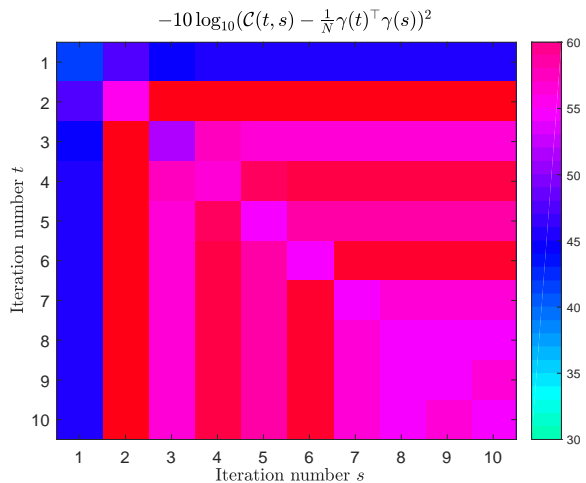,width=1.04\columnwidth}
	\caption{Discrepancy between theory and simulations for the 
	covariance in case of a $2$-layer Hopfield model with $N/N_1=1$ and $N/N_2=2$, $N=10^{4}$ and $\beta=0.25$. }\label{fig1}
\end{figure}

\begin{figure}
	\epsfig{file=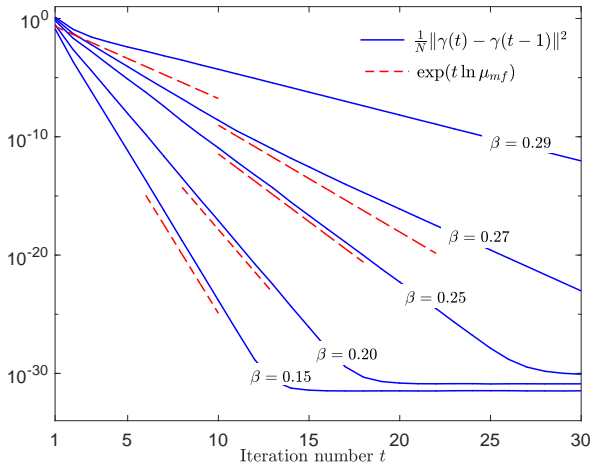,width=1.03\columnwidth}
	\caption{Asymptotics of the algorithm for $2$-layer Hopfield model with $N/N_1=1$ and $N/N_2=2$, $N=10^{4}$ and $h=1$. The inverse temperature $\beta=0.35$ gives the AT line. The flat lines around $10^{-30}$ are the consequence of the machine precision of the computer which was used. }\label{fig2}
\end{figure}

In this section, we compare our analytical results with simulations of the algorithm 
for large random matrices.
We note from \eqref{deltaC} that the dynamics \eqref{dynamics} in the thermodynamic limit 
is described in terms of the limiting “covariance” expression
\begin{equation}
\mathcal C(t,s)=\lim_{N\to\infty}\frac{1}{N} \mathbb E[\matr \gamma(t)^\top \matr \gamma(s)]. \label{covdef}
\end{equation}
whose explicit analytical expression is given in \eqref{covfield}. In Figure~\ref{fig1} 
we illustrate how well $\mathcal C(t, s)$ predicts $\frac{1}{N}\matr \gamma(t)^\top \matr \gamma(s)$ for a $2$-layer Hopfield model with a single realization of the dynamics. Note that we assume the large-system limit $N\to\infty$ is taken before the long-time limit $t\to\infty$. Thus, we get excellent agreement between theoretical predictions and simulations on single instances for finite-time properties of large systems. The discrepancy between theory and simulations for large times increases
as the model parameters approach the AT line. This can be seen in Figure~\ref{fig2},
close to the AT line (i.e. $\beta=2.29$) where \eqref{res2} fails to predict the exponential decay for a single-realization of dynamics. In fact the dynamics has started to lose its self-averaging properties in this regime, i.e. the dynamics for different realisations of the random matrix and random
initial conditions yield remarkably different asymptotic convergence behavior, see also in Figure~\ref{fig4}. In Figure~\ref{fig1} and Figure \ref{fig2}, we compute the necessary quantities $\rm R$ and $\rm R'$ by the algorithmic approach in Section~\ref{algorc}.
 
\begin{figure}
	\epsfig{file=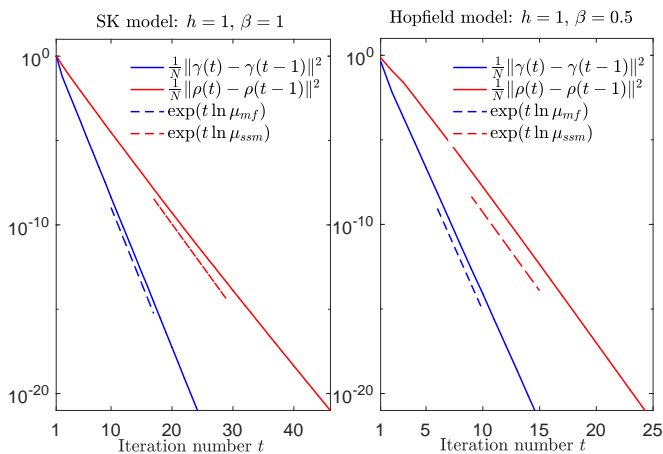,width=1.03\columnwidth}
	\caption{Comparison of the convergence rates with SSM algorithms for the SK and Hopfield models. $N=10^4$.}\label{fig3}
\end{figure}

\begin{figure}
	\epsfig{file=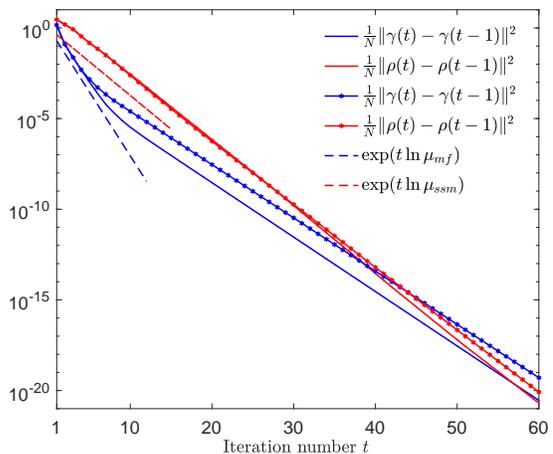,width=1.03\columnwidth}
	\caption{Asymptotics of the algorithm for Hopfield model with $h=1$, $\beta=0.75$ and $N=10^4$. We show the results of two different realizations (reds and blues) for the random 
	matrix and initial conditions in the dynamics \eqref{dynamics} and \eqref{amp}. The inverse temperature $\beta=1.08$ gives the AT-line. }\label{fig4}
\end{figure}

Figure~\ref{fig3} refers to the SK and (1-layer) Hopfield models  
where we compare the convergence rates of the dynamics \eqref{dynamics} with the SSM dynamics for a single-realization. The simulation results confirm the thermodynamic bound \eqref{iden}. However, as we illustrate in Figure~\ref{fig4} (for the Hopfield model)
that when the model parameters approach the AT line, after few iteration steps the dynamics \eqref{dynamics} shows strong fluctuations with respect to different realizations of 
while the SSM dynamics shows negligible fluctuations.  

\begin{figure}
	\epsfig{file=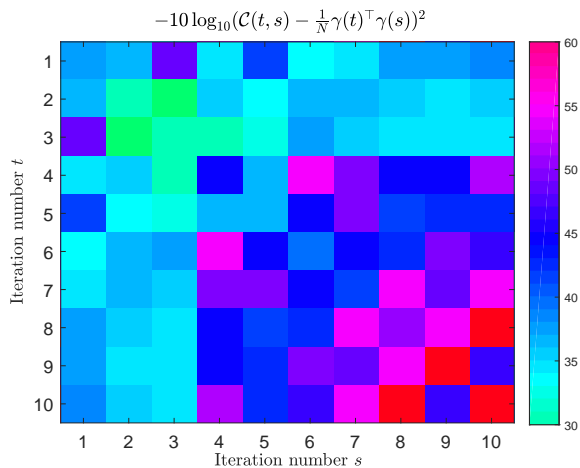,width=1.03\columnwidth}
	\caption{Discrepancy between theory and simulations for the 
	covariance in case of randomly-signed Hadamard model with $N=2^{13}$, $h=2$, $\alpha=1/2$ and $\beta=4$.}\label{fig5}
\end{figure}

\begin{figure}
\epsfig{file=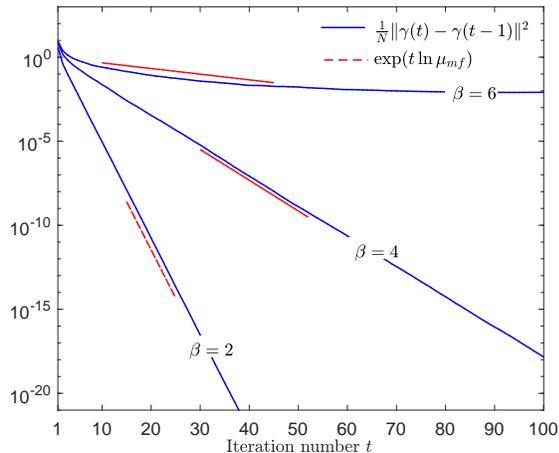,width=1.03\columnwidth}
\caption{Asymptotics of the algorithm for randomly-signed Hadamard model with $N=2^{13}$, $h=2$, $\alpha=1/2$. The inverse temperature $\beta=6.8$ gives the AT line.}\label{fig6}
\end{figure}

So far we have shown examples of the performance of the algorithm on matrices
drawn at random from spherical invariant distributions. On the other hand, all necessary 
parameters (R-transform and its derivative) of the algorithm can be computed from a given matrix without explicit knowledge of the distribution. The same parameters appear in the analytical 
results of the theory and depend only on the (limiting) spectrum of the matrix.  What happens if we run the algorithm on a matrix which is not a typical realization from a spherical invariant random matrix ensemble? To get a first impression on the robustness of our results we  compare the theoretical results (obtained from the spectrum) with simulations based on a model
with \emph{randomly-signed Hadamard} matrices. To understand this model we first consider the
(spherical symmetric) random orthogonal model discussed by Parisi and Potters \cite{Parisi} (i.e. the 1-layer random-orthogonal model with $\rho_0=1$). The coupling matrix can be written in the form
\begin{equation}
\matr J=\beta\matr O^\top\matr D_{\rho}\matr O
\end{equation}
where $\matr O$ is a Haar matrix and $\matr D_{\rho}={\rm diag}(d_1,\cdots,d_N)$ has 
random binary elements $d_i=\mp 1$ with $\vert\{d_i=1\}\vert=\rho N$.
Motivated by a recent study \cite{Greg} in random matrix theory we now introduce a non-rotation invariant coupling matrix ensemble ---  called \emph{randomly-signed Hadamard}
by replacing the random Haar matrix by a different random orthogonal matrix $\tilde{\matr O}$:
\begin{equation}
\matr J=\beta\tilde{\matr O}^\top\matr D_{\rho}\tilde{\matr O}~~ \text{with}~~\tilde{\matr O}\doteq \frac{1}{\sqrt{N}}\matr H_{N}\matr Z. \label{rsh}
\end{equation}
Here, $\matr Z$ is an $N\times N$ diagonal matrix whose diagonal entries are independent and composed of binary $\mp1$ random variables with equal probabilities and $\matr H_N$ is the $N\times N$ Hadamard matrix which is {\em deterministically} constructed \cite{hadamard1893resolution}
from the recursion 
\begin{align}
\matr H_{2^k}=\left[\begin{array}{cc}
\matr H_{2^{k-1}}& \matr H_{2^{k-1}}\\
\matr H_{2^{k-1}}&-\matr H_{2^{k-1}}
\end{array}\right], ~~ k\geq 1
\end{align}
with $\matr H_{1}\doteq1$. Note that $\tilde{\matr O}$ is an orthogonal matrix with the {\em binary entries} $\tilde O_{ij}=\mp \frac{1}{\sqrt{N}}$. Both coupling matrices have the same spectra and thus the same
R-transform which can be easily obtained as 
\begin{equation}
{\rm R}(\omega)=\frac{(\beta \omega-\rho)+\sqrt{(\beta \omega-\rho)^2+4\rho^2\beta\omega}}{2\rho\omega}-\beta.
\end{equation}
In Figure~\ref{fig5}~and Figure~\ref{fig6}
we illustrate how well our theoretical results describe the dynamics \eqref{dynamics} when the coupling matrix $\matr J$ is modeled by a randomly-signed Hadamard ensemble. The simulation results suggest that our theoretical results can be valid for larger classes of coupling matrices than the rotation invariant matrix ensembles. 
\section{Summary and Outlook}
In this paper we have introduced an iterative algorithm for solving the TAP equations of Ising models with arbitrary rotation invariant coupling matrices. We have shown that the algorithm is convergent in the thermodynamic limit when the AT-line criteria is fulfilled. We have also obtained a compact analytical expression for the rate of convergence. The analytical results are supported by numerical simulations of large systems on single instances of random matrices. Preliminary simulations for random matrices which are not generated from rotation invariant ensembles but show similar weak dependencies between matrix elements indicate that our theory may extend to larger classes of matrices. Preliminary analytical calculations suggest that some of our results on convergence can be derived under the weaker property of {\em asymptotic freeness} \cite{mingo2017free} (which is fulfilled  by rotation invariant matrix ensembles). Such extensions of our approach would also be important for applications to TAP-style mean field equations 
for statistical inference in probabilistic models. A first step in this direction would be to consider Ising models with non-constant or random external fields. Such problem setting plays a role when  the probability distribution of the spins is conditioned on observed spin variables. This is a relevant problem in machine learning e.g. in the training  of restricted Boltzmann machines~\cite{Tramel18}. More complex extensions would be to the important case of models with continuous  random variables. Of course, the latter case
would require new and nontrivial generalizations of our basic algorithm. These generalizations
might be motivated e.g. by the so-called {\em expectation propagation algorithms} \cite{Minka1,OW5} which are 
frequently used in machine learning.  A different direction of research is in the optimization of algorithms. It will be interesting to investigate which properties makes our new method converge faster than the previously defined SSM algorithms. On the other hand it will also be important to 
get a better understanding of finite size fluctuations, because these will influence the robustness of convergence, when the dynamics is close to the AT instability.

\section*{Acknowledgment}
We are grateful to funding from the BMBF (German ministry of education and research) joint project {\bf01 IS 18037 A} : BZML- Berlin Center for Machine Learning.

\appendix

\section{Vanishing of memory terms}\label{vanishmemory}
Using the definition (\ref{resp}) and
differentiating (\ref{gam2}) with respect to $\phi(\tau)$ for $\tau < t$, we obtain
\begin{eqnarray}
\mathcal G(t,\tau) =
\mathbb E\left[f'(\gamma(t-1))\frac{\partial \gamma(t-1)}{\partial \phi(\tau)}\right] 
\nonumber\\
= \sum_{\tau < s < t} \hat{\mathcal{G}}(t -1,s) \mathbb E\left[f'(\gamma(t-1))\frac{\partial \gamma(t-1)}{\partial \phi(\tau)}\right] 
\nonumber
\\
+ \mathbb E\left[f'(\gamma(t-1)) \right]\delta_{t-1,\tau}
 \label{resonse_it}
\end{eqnarray}
for $t=1,2,\ldots,T$. 
From equations (\ref{resonse_it}) and the definitions of $\hat{\mathcal{G}}$ and
$\mathcal{C}$, one can show that these quantities are uniquely defined recursively 
in time as expectations over the stochastic process $\phi$ and the random initial conditions. 
Hence, if we can show 
that the trivial solution $\mathcal{G} = \hat{\mathcal{G}}=\matr 0$ fulfills these equations, we will 
be done. Obviously, this solution fulfills the relation between $\mathcal{G}$ and $\hat{\mathcal{G}}$
eq. (\ref{Ghat}). On the other hand, if $\hat{\mathcal{G}}= \matr 0$,  we obtain
$\mathcal{G} =\matr0$ from (\ref{resonse_it}), {\em provided} we can show
that 
\begin{equation}
\mathbb E\left[f'(\gamma(t-1)) \right] = 0 \label{vanish_fprime}
\end{equation}
is consistent with these assumptions. The proof of this fact and the resulting form of the covariance matrix \eqref{covfield} is deferred to Appendix~\ref{dcovfield}. 

\subsection{The covariance matrix \eqref{covfield}}\label{dcovfield}
By using the definition of $\tilde{\mathcal C}$ in (\ref{tilde_cov}) and using (\ref{gam1}) we obtain the recursion
\begin{align}
\tilde{\mathcal{C}}(t,s) &  = \frac{1}{\chi^2}{\mathbb E}[{\rm Th}(\gamma(t-1)){\rm Th}(\gamma(s-1))]  + 
\nonumber \\
& 
\left( 1 
- \frac{1}{\chi}{\mathbb E}[{\rm Th'}(\gamma(t-1))]
- \frac{1}{\chi}{\mathbb E}[{\rm Th'}(\gamma(s-1))]
\right) \times \nonumber \\
& \times \mathcal{C}(t-1,s-1).
\label{recurs_tilde}
\end{align}
Note, that this result is also true for $s=t=1$, if we set 
\begin{equation}
\mathcal{C}(0,0) \doteq \mathbb E[\gamma^2(0)] = (1-\chi){\rm R}'.
\end{equation}

Recall that by the AT line condition \eqref{cond2}, the equation \eqref{chi} has a unique solution for $\chi$. Then, for the variance terms $\mathcal C(t,t)$, we can show by induction that 
\begin{equation}
\mathcal C(s,s)=(1-\chi){\rm R}' \qquad s > 0 \label{seq}
\end{equation}
is independent of time. Specifically, assuming that (\ref{seq}) is true for time $s= t-1$.
Hence, we have
\begin{equation}
\frac{1}{\chi}{\mathbb E}[{\rm Th'}(\gamma(t-1))] = 1\label{fvar}
\end{equation}
which in turn is equivalent to (\ref{vanish_fprime}). We then get at the next time step
\begin{align}
\mathcal C(t,t) &=\frac{\chi^2 {\rm R}'}{1-\chi^2 {\rm R}'}\left[\frac{1-\chi}{\chi^2} -\mathcal C(t-1,t-1)\right]\label{arg1}\\
&=(1-\chi){\rm R}' . \label{iden2}
\end{align}
Moreover, plugging \eqref{fvar} into \eqref{recurs_tilde} we have
\begin{equation}
\tilde{\mathcal{C}}(t,s)=\frac{1}{\chi^2}{\mathbb E}[{\rm Th}(\gamma(t-1)){\rm Th}(\gamma(s-1))]-\mathcal C(t-1,s-1)
\end{equation}
which completes the derivation of \eqref{covfield}.

\section{Derivations of \eqref{res1}, \eqref{res2} and \eqref{res3}} \label{dres13}
\begin{remark}[A useful remark]\label{useful}
	Let $\gamma_1$ and $\gamma_2$ be Gaussian random variables and be identically distributed. Furthermore, let $\mathcal C_{12}$ stand for the covariance between $\gamma_1$ and $\gamma_2$ such that $\mathcal C_{12}>0$. Moreover, let the function $f$ have derivatives of all orders in~$\RR$. Then, the covariance between the random variables $f(\gamma_1)$ and $f(\gamma_2)$ is positive, too. 
\end{remark}
The derivation of Remark~\ref{useful} is located in Appendix~\ref{duseful}.
\subsection{Derivation of \eqref{res1}}\label{dres1}
Using the representation of the Gaussian density in terms of the characteristic function (see the argument of \eqref{Gcar}) one can show that
\begin{equation}
g'(x)={\mathbb E[{\rm Th}'(\gamma_1){\rm Th}'(\gamma_2)]} -\chi^2\label{derg}
\end{equation}
where $g'$ denotes the derivative of the function $g$ in \eqref{kfunc}. Note that $\chi=\mathbb E[{\rm Th}'(\gamma_1)]$. Thus, $g'(x)$ equals the covariance between the random variables ${\rm Th'(\gamma_1)}$ and ${\rm Th}'(\gamma_2)$. Thereby, from the Remark~\ref{useful} it turns out that 
\begin{equation}
g'(x)>0, \quad x>0. \label{positivity}
\end{equation}

To prove \eqref{res1} it is sufficient to show 
\begin{equation}
\mathcal C(t+1,s+1)>\mathcal C(t,s),\quad \forall t,s.\label{strictp}
\end{equation}
Since $g(x)$ is strictly increasing on the positive axis, assuming \eqref{strictp} holds and $\mathcal C(t,s)>0$ we have
\begin{align}
\mathcal C(t+2,s+2)&=\frac{{\rm R}'}{1-\chi^2{\rm R}'}g(\mathcal C(t+1,s+1))\\
&>\frac{{\rm R}'}{1-\chi^2{\rm R}'}g(\mathcal C(t,s))\\
&=\mathcal C(t+1,s+1).
\end{align}
Thus, we complete the derivation by showing the inequality $\mathcal C(t+1,1)>\mathcal C(t,0)$, $\forall t$. This follows from the fact that $g(0)~>~0$ and \eqref{s1}, i.e. $\mathcal C(t,0)=0$.

\subsection{Derivations of \eqref{res2}}\label{edecay}
From \eqref{covfield} and \eqref{npt} we can write $\Delta(t+1,s+1)$ in the form
\begin{equation}
\Delta(t+1,s+1)=d(\Delta(t,s))
\end{equation}
for an appropriately defined function $d$. In particular, we note from \eqref{npt} that 
\begin{align}
\frac{\partial \Delta(t+1,s+1)}{\partial\Delta(t,s)}&=\frac{\partial \mathcal C(t+1,s+1)}{\partial \mathcal C(t,s)}\\
&=\frac{{\rm R}'}{1-\chi^2{\rm R}'}g'(\mathcal C(t,s))\label{dg}
\end{align}
where the function $g'$ is as defined in \eqref{derg}. Since $g'$ has derivative in all order (see Remark~\ref{guseful}), for sufficiently large $t$ and $s$ we can expand $\Delta(t+1,s+1)$ around $0$ as
\begin{equation}
\Delta(t+1,s+1)=d'(0)\Delta(t,s)+O(\Delta(t,s)^2)
\end{equation}
with noting that $d(0)=0$. Hence, we get
\begin{align}
	\lim_{t,s\to\infty}\frac{\Delta(t+1,s+1)}{\Delta(t,s)}&=\lim_{t,s\to\infty}(d'(0)+O(\Delta(t,s)))\nonumber \\
	&=d'(0).
\end{align}
We now recall that the function $g'(x)$ equals to the covariance between the random variables ${\rm Th'(\gamma_1)}$ and ${\rm Th}'(\gamma_2)$ where $\gamma_1$ and $\gamma_2$ are jointly zero-mean Gaussian random variables with equal variance $(1-\chi){\rm R}'$ and covariance $x$. Thus, we have
\begin{align}
d'(0)&=\frac{{\rm R}'}{1-\chi^2{\rm R}'}g'((1-\chi){\rm R}')\\
&=\frac{{\rm R}'}{1-\chi^2{\rm R}'}(\mathbb E[({\rm Th}'(\gamma_1))^2] -\chi^2)\\
&= 1-\frac{1-\eta{\rm R}'}{1-\chi^2{\rm R}'}.
\end{align}
This completes the derivation. 

\subsection{Derivations of \eqref{res3}}
For short let
\begin{align}
\Delta(t)&\doteq \lim_{s\to \infty}\Delta(t,s)=2(1-\chi){\rm R}'-2\mathcal C(t).
\end{align}
where we have defined
\begin{equation}
\mathcal C(t)\doteq\lim_{s\to\infty}\mathcal C(t,s).\label{rate}
\end{equation}
We are interested in the rate of the asymptotic decay 
\begin{equation}
\Delta(t) \simeq e^{t\kappa}\qquad t\to\infty.
\end{equation}
The rate $\kappa$ can be computed as 
\begin{equation}
\kappa =\lim_{t\to\infty} \ln \frac{\partial \Delta(t+1)}{\partial\Delta(t)}.\label{kappa}
\end{equation}
The derivative is obtained from the recursion of $\mathcal C(t)$ 
\begin{equation}
\mathcal C(t+1)=\frac{g(\mathcal C(t))}{1/{\rm R}'-\chi^2}
\end{equation}
where the function $g$ is defined in \eqref{kfunc}. Then, we have
\begin{align}
\frac{\partial \Delta(t+1)}{\partial\Delta(t)}
& =\frac{\partial \mathcal C(t+1)}{\partial \mathcal C(t)}\label{s}\\
&=\frac{g'(\mathcal C(t))}{1/{\rm R}'-\chi^2}\label{f}\\
&=\frac{\eta(\mathcal C(t))-\chi^2}{1/{\rm R}'-\chi^2}
\end{align}
Here, the function $g'$ is as in \eqref{derg} and for convenience we introduce the function 
\begin{equation}
\eta(x)\doteq g'(x)+\chi^2.
\end{equation}
Following the steps of \ref{dres1} we get $\mathcal C(t+1)>\mathcal C(t)$, $\forall t$. Moreover, by using Remark~\ref{useful} we have that $\eta(x)$ is a strictly increasing function on the positive axis. Thus, we have
\begin{equation}
\eta (\mathcal C(t))<\eta(\mathcal C(t+1))<\eta,\quad \forall t.
\end{equation}
Hence, we have
\begin{equation}
\lim_{t\to \infty}\eta (\mathcal C(t))=\eta
\end{equation}
which completes the derivation.

\subsection{Derivation of Remark~\ref{useful}}\label{duseful}
Remark~\ref{useful} evidently follows from the following general result.
\begin{remark}\label{guseful}
	Let $\gamma_1$ and $\gamma_2$ be Gaussian random variables with means $\mu_1$ and $\mu_2$ and variances $\sigma_1^2$ and $\sigma_2^2$, respectively. Moreover, let $\mathcal C_{12}$ stand for the covariance between $\gamma_1$ and~$\gamma_2$. Also, let the function $f$ have derivatives of all orders in $\RR$. Then, the covariance between the random variables $f(\gamma_1)$ and $f(\gamma_2)$ is given by
	\begin{equation}
	\sum_{n=1}^\infty\frac{\mathcal C_{12}^n}{n!}\mathbb E\left[\frac{{\rm d}^nf(\gamma_1)}{{\rm d}\gamma_1^n}\right]\mathbb E\left[\frac{{\rm d}^nf(\gamma_2)}{{\rm d}\gamma_2^n}\right].
	\end{equation}
\end{remark}
The derivation of Remark~\ref{guseful} is based on the representation of the Gaussian density in terms of the characteristic function:
\begin{align}
&\mathbb E[f(\gamma_1) f(\gamma_2)]=\frac{1}{(2\pi)^2}\int {\rm d}\gamma_1{\rm d}\gamma_2{\rm d}k_1{\rm d}k_2\;f(\gamma_1) f(\gamma_2)\times \nonumber \\ &\times e^{{\rm i}k_1(\mu_1-\gamma_1)+{\rm i}k_2( \mu_2-\gamma_2) -\frac{1}{2}(\sigma_1^2k_1^2+\sigma_2^2k_2^2)}e^{-\mathcal C_{12}k_1k_2}\label{Gcar} \\&=\mathbb E[f(\gamma_1)]\mathbb E[f(\gamma_2)]+\sum_{n=1}^\infty\frac{\mathcal C_{12}^n}{n!}\frac{1}{(2\pi)^2}\int {\rm d}\gamma_1{\rm d}\gamma_2{\rm d}k_1{\rm d}k_2\;\times \nonumber \\
&(-k_1k_2)^n{f}(\gamma_1) {f}(\gamma_2)e^{{\rm i}k_1(\mu_1-\gamma_1)+{\rm i}k_2( \mu_2-\gamma_2) -\frac{1}{2}(\sigma_1^2k_1^2+\sigma_2^2k_2^2)}\label{taylor}\\
&=\mathbb E[f(\gamma_1)]\mathbb E[f(\gamma_2)]+\sum_{n=1}^\infty\frac{\mathcal C_{12}^n}{n!}\mathbb E\left[\frac{{\rm d}^nf(\gamma_1)}{{\rm d}\gamma_1^n}\right]{\mathbb E\left[\frac{{\rm d}^nf(\gamma_2)}{{\rm d}\phi_2^n}\right]} \label{derk1}
\end{align}
where in \eqref{taylor} we use the Taylor expansion representation of $e^{-\mathcal C_{12}k_1k_2}$ and in \eqref{derk1} we represent $(-k_1k_2)^n$ by means of the derivatives $\frac{\partial^n}{\partial\gamma_1^n}\frac{\partial^n}{\partial\gamma_2^n}$. 

\section{Derivations of \eqref{delta_ssm} and \eqref{rho_con}}\label{amp_con}
Let us introduce
\begin{align}
\mathcal C_{\rho}(t,s)&\doteq \lim_{N\to \infty}\frac{1}{N}\mathbb E[\matr \rho(t)^\top \matr \rho(s)]
\end{align}
where the expectation is taken over the random matrix ensemble of $\matr J$ and the random initialization $\matr\rho(0)$. We will next show from the results of \cite{Opper16} that for both the SK and Hopfield models we have
\begin{equation}
\mathcal C_{\rho}(t,s)={\rm R}'\mathbb E\left[{\rm Th}(\rho(t-1)){\rm Th}(\rho(s-1))\right]\label{covpsi}
\end{equation}
where the expectation is taken over the random variables $\rho(t)$ and $\rho(s)$ which are jointly zero-mean Gaussian with equal variances $(1-\chi){\rm R}'$ and covariance $\mathcal C_{\rho}(t,s)$. Also we note from \eqref{covpsi} that
\begin{equation}
\frac{\partial \mathcal C_{\rho}(t+1,s+1)}{\partial\mathcal C_{\rho}(t,s)}={\rm R}'\mathbb E\left[{\rm Th}'(\rho(t)){\rm Th}'(\rho(s))\right].
\end{equation}
Then, by following the steps of Appendix~\ref{dres1} and~\ref{edecay} one can obtain \eqref{delta_ssm} and \eqref{rho_con}, respectively. 

To read off the result \eqref{covpsi} from \cite{Opper16} for the SK model, we first write the SSM updates \cite[Eq. (35)--(38)]{Opper16} for the SK model as \cite{Opper16}
\begin{subequations}
	\label{ssm_sk}
	\begin{align}
	\matr m(t)&={\rm Th}(\matr \rho(t-1))\\
	\matr \rho(t)&=\matr J\matr m(t)-\beta^2\chi(t)\matr m(t-1)
	\end{align}
\end{subequations} 
where $\matr m(0)=\matr 0$ and we have defined
\begin{equation}
\chi(t)\doteq 1-\lim_{N\to \infty}\frac{1}{N}\mathbb E[\matr m(t)^\top \matr m(t)].\label{chit}
\end{equation}
In this case, the two-time covariance $\mathcal C_\rho(t,s)$ is read off from \cite[Eq. (60)]{Opper16} as
\begin{equation}
\mathcal C_{\rho}(t,s)=\beta^2\mathbb E\left[{\rm Th}(\rho(t-1)){\rm Th}(\rho(s-1))\right]\label{covamp_sk}
\end{equation}
where the expectation is taken over the random variables $\rho(t)$ and $\rho(s)$ which are jointly zero-mean Gaussian with covariance $\mathcal C_{\rho}(t,s)$. Moreover, we have 
\begin{equation}
\chi(t)=\mathbb E\left[{\rm Th}'(\sqrt{\mathcal C_{\rho}(t-1,t-1)}u)^2\right]. \label{tchi_sk}
\end{equation}
From the initialization $\matr\rho(0)=\sqrt{(1-\chi)\beta^2}\matr u$  we have $\mathcal C_{\rho}(0,0)=(1-\chi)\beta^2$. Then, it follows inductively from \eqref{covamp_sk} and \eqref{tchi_sk} that $\mathcal C_{\rho}(t,t)=(1-\chi)\beta^2$ so that we have
\begin{equation}
\chi(t)=\chi, \quad \forall t.
\end{equation}
Using this result in \eqref{ssm_sk} and \eqref{covamp_sk} leads to \eqref{amp_sk} and \eqref{covpsi}, respectively.  

To read off the result \eqref{covpsi} from \cite{Opper16} for the Hopfield model, we follow the definition of the SSM construction \cite[Section~4]{Opper16} and define the following dynamics  \begin{subequations}
\label{amporg}
\begin{align}
\matr m(t)&={\rm Th}(\matr \rho(t-1))\\
\matr \phi(t)&=\matr J\matr m(t)-\frac{\chi(t)}{\chi(t-1)}{\rm R}(\chi(t-1))\matr m(t-1)\\
\matr \rho(t)&=[1+\frac 1 \beta{\rm R}(\chi (t))]\matr \phi(t)-\frac 1 \beta{\rm R}(\chi(t))\matr \rho(t-1)\label{psit}
\end{align}
\end{subequations} 
where $\matr m(0)=\matr 0$, $\chi(t)$ is as in \eqref{chit}. The R-transform for the Hopfield model
is given by ${\rm R}(\omega)=\frac{\alpha_1\beta^2\omega}{1-\alpha_1\beta\omega}$. Basically, $\matr\phi(t)$ and $\matr\rho(t)$ play the roles of $\matr\phi(t)$ and $\sum_{\tau=0}^{t}\mathcal A(t,\tau)\matr \phi(\tau)$ in \cite{Opper16}, respectively. Thus, we have from \cite[Eq. (60)]{Opper16}
\begin{align}
\mathcal C_{\rho}(t,s)&=\frac{{\rm R}(\chi(t)){\rm R}(\chi(s))}{\beta^2\alpha_1\chi(t)\chi(s)}\mathbb E\left[{\rm Th}(\rho(t-1)){\rm Th}(\rho(s-1))\right]\label{covamp}
\end{align}
where $\chi(t)$ reads as in \eqref{tchi_sk}. Again, from the initialization $\matr\rho(0)=\sqrt{(1-\chi){\rm R}'}\matr u$ we have $\mathcal C_{\rho}(0,0)=(1-\chi){\rm R}'$. It then follows inductively that $\mathcal C_{\rho}(t,t)=(1-\chi){\rm R}'$ so that we have
\begin{equation}
\chi(t)=\chi, \quad \forall t.
\end{equation}
Using this result in \eqref{ssm_sk} and \eqref{covamp_sk} lead to \eqref{amp} and \eqref{covpsi}, respectively.

\section{Derivation of \eqref{limitsk}}\label{dlimitsk}
We first note the scaling property of the R-transform \cite{ralfc}
\begin{equation}
{\rm R}_{c\matr J}(\omega)=c{\rm R}_{\matr J}(c\omega)\label{scaler}
\end{equation} 
where ${\rm R}_{(\cdot)}$ is the R-transform of the limiting spectral distribution of the matrix given in subscript. Second, by an appropriate reformulation of the result \cite[Corollary 1.14]{nica1} --- known as free compression of random matrices --- we write
\begin{equation}
{\rm R}_{\matr P^{N_M}_{\alpha}\matr Y(\matr P^{N_M}_{\alpha})^\top}(\omega)={\rm R}_{\matr Y}(\alpha\omega)\label{comp}.
\end{equation}
Then, by using \eqref{scaler} and \eqref{comp} we get
\begin{align}
{\rm R}_{\frac{1}{\sqrt{\alpha}}\matr J}(\omega)&=\frac{1}{\sqrt{\alpha}}{\rm R}_{\matr Y}\left({\sqrt{\alpha}}\omega\right)\\
&=\frac{1}{\sqrt{\alpha}}\sum_{n=1}^{\infty}c_n{\alpha}^{\frac{n-1}{2}}\omega ^{n-1}.
\end{align}
Here, $c_n$ denotes the $n$th free cumulant of the limiting spectral distribution of $\matr Y$ such that $c_1$ and $c_2$ are the mean and variance of the distribution, respectively, i.e. $c_1=0$ and $c_2=\sigma_{\matr Y}^2$. Thus, we get
\begin{equation}
\lim_{\alpha\to 0}{\rm R}_{\frac{1}{\sqrt{\alpha}}\matr J}(\omega)=\sigma_{\matr Y}^2\omega
\end{equation}
which is nothing but the R-transform of a Wigner semicircle distribution \cite{Hiai}. This completes the derivation. 
\section{Additivity of $\lim_{N\to\infty}\frac{1}{N}{\rm tr}(\matr J^2)$}\label{addsigma}
Let $\matr J$ be modeled as in \eqref{Mlayer} and let all $N_m\times N_{m-1}$ matrices $\matr X_m$ in \eqref{prod} be \emph{asymptotically free} of each others. Moreover, for convenience we assume the normalization
\begin{equation}
\lim_{N\to\infty}\frac 1 N{\rm  tr}(\matr X_m^\top\matr X_m)=1,\quad \forall m.
\end{equation}
Then, we have 
\begin{equation}
\sigma^2_{\matr J}=\beta^2\sum_{m\leq M}\alpha_m\sigma_{\matr X_m^\top\matr X_m}^2\label{nnlemma}
\end{equation}
where $\alpha_m\doteq \frac{N_0}{N_m}$ and $\sigma^2_{(\cdot)}$ denotes the variance of the limiting spectral distribution of the matrix given in subscript. Here, for the definition of \emph{asymptotic freeness} of random matrices we refer the reader to \cite{ralfc,Tulino}. Nevertheless, we note that for the multi-layer Hopfield and multi-layer random orthogonal models the aforementioned asymptotic freeness condition holds (almost surely)~\cite{Hiai}. 

The result \eqref{nnlemma} follows from the properties of the (Voiculescu) S-transform \cite{VoicuS}. Specifically, let ${\rm F}$ be a probability distribution on real line with a non-zero mean and let ${\rm R}$ be its R-transform. The S-transform of ${\rm F}$ is defined by the functional inverse of $\tilde{\rm R}(\omega)\doteq \omega{\rm R}(\omega)$ as \cite{hager3}
\begin{equation}
{\rm S}(s)\doteq \frac{1}{s}\tilde{\rm R}^{-1}(s). \label{S-trans}
\end{equation}
In particular, for $c_1$ and $c_2$ denoting the mean and variance of ${\rm F}$, respectively, we have \cite[Page 207]{larsen}
\begin{equation}
{\rm S}(0)=\frac{1}{c_1}\quad  \text{and} \quad {\rm S}'(0)=\frac{c_2}{c_1^3} \label{smean}.
\end{equation}
Moreover, for asymptotically free matrices, say $\matr A=\matr A^\top$ and $\matr B=\matr B^\top$, we have ${\rm S}_{\matr A\matr B}(s)={\rm S}_{\matr A}(s){\rm S}_{\matr B}(s)$ \cite{VoicuS} where ${\rm S}_{(\cdot)}$ denotes the S-transform of the limiting spectral distribution of the matrix given in subscript. Combining this property of the S-transform with \cite[Lemma 4.3]{debbah} one can show that
\begin{equation}
{\rm S}_{\matr X^\top\matr X}(s)=\prod_{m\leq M}{\rm S}_{\matr X_m^\top\matr X_m}(\alpha_ms).\label{sdecom}
\end{equation}
Note that $\sigma_{\matr J}^2=\beta^2\sigma_{\matr X^\top\matr X}^2$. Then, \eqref{nnlemma} follows easily by invoking the properties of the S-transform \eqref{smean} for \eqref{sdecom}. 

\section{Derivation of \eqref{rmodel2}}\label{drmodel2}
From \cite[Theorem 2]{ralfa} we write the S-transform (see the definition in \eqref{S-trans}) of the limiting spectral distribution of $\matr X^\top\matr X$ as
\begin{equation}
{\rm S}_{\matr X^\top\matr X}(s)=\frac{1}{\alpha_1s+1}\frac{1}{\alpha_2s+1}\label{s2hopfield}.
\end{equation}
Then, from \eqref{S-trans} we get the R-transform of the limiting spectral distribution of $\matr X^\top\matr X$ 
\begin{align}
{\rm R}_{\matr X^\top\matr X}(\omega)&=\frac{\frac{1}{ \omega}-\alpha_1-\alpha_2\mp \sqrt{(\frac{1}{ \omega}-\alpha_1-\alpha_2)^2-4\alpha_1\alpha_2}}{2\alpha_1\alpha_2\omega}\label{r2hopfield}.
\end{align}
Moreover, with the scaling property of the R-transform (see \eqref{scaler}) we have
\begin{align}
{\rm R}_{\matr J}(\omega)&=\beta{\rm R}_{\matr X^\top\matr X}(\beta\omega)-\beta{\rm R}_{\matr X^\top\matr X}(0)\\
&=\beta{\rm R}_{\matr X^\top\matr X}(\beta\omega)-\beta.
\end{align}
From \eqref{r2hopfield} we have two solutions of ${\rm R}_\matr J(\omega)$. Only one fulfills the property ${\rm R}_\matr J'(\omega)>0$. This completes the derivation.


\bibliography{mybib}

\end{document}